\documentclass[english,twocolumn]{revtex4-1}

\usepackage[pdftex]{graphicx}
\usepackage{caption}
\usepackage{subcaption}
\usepackage{savesym}
\usepackage{listings}
\usepackage{float}
\usepackage{epsfig}
\floatstyle{boxed}
\restylefloat{figure}
\usepackage{amsthm}
\usepackage{babel}
\usepackage[intlimits]{amsmath}
\savesymbol{iint}
\usepackage{txfonts}
\restoresymbol{TXF}{iint}
\usepackage{amssymb}
\usepackage{comment}
\usepackage{color}
\usepackage{cancel}

\def\be{\begin{equation}}
\def\ee{\end{equation}}
\def\ba{\begin{eqnarray}}
\def\ea{\end{eqnarray}}
\def\go{\mathrel{\raise.3ex\hbox{$>$}\mkern-14mu
             \lower0.6ex\hbox{$\sim$}}}
\def\lo{\mathrel{\raise.3ex\hbox{$<$}\mkern-14mu
             \lower0.6ex\hbox{$\sim$}}}

\usepackage{babel}
\begin{document}

\title{Traversable Achronal Retrograde Domains In Spacetime}

\author{Benjamin K. Tippett}\email{btippett@mail.ubc.ca} \affiliation{University of British Columbia, Okanagan, 3333 University Way, Kelowna BC V1V 1V7, Canada}

\author{David Tsang}\email{dtsang@physics.mcgill.ca} \affiliation{Physics Department, McGill University, Montreal, QC, H3A 2T8, Canada}

\begin{abstract}
There are many spacetime geometries in general relativity which contain closed timelike curves. A layperson might say that \emph{retrograde time travel}  is possible in such spacetimes. To date no one has discovered a spacetime geometry which emulates what a layperson would describe as a \emph{time machine}.  The purpose of this paper is to propose such a space-time geometry. 

In our geometry,  a bubble of curvature travels along a closed trajectory. The inside of the bubble is Rindler spacetime, and the exterior  is Minkowski spacetime. Accelerating observers inside of the bubble travel along closed timelike curves. The walls of the bubble are generated with matter which violates the classical energy conditions. We refer to such a bubble as a Traversable Achronal Retrograde Domain In Spacetime.   \end{abstract}
\maketitle

\section{Introduction}
\subsection{Exotic  spacetimes}

How would one go about building a \emph{time machine}? Let us begin with considering exactly what one might mean by ``time machine?" H.G. Wells (and his successors) might describe an apparatus which could convey people ``backwards in time". That is to say, convey them from their current location in spacetime to a point within their own causal past.  If we describe spacetime as a river-bed, and the passage of time as our unrelenting flow along our collective worldlines towards our future; a time machine would carry us back up-stream. Let us describe such motion as \emph{retrograde time travel}.
 
 In the context of general relativity, just as in the river, it is the underlying geometry which determines where the worldlines can flow. In general, it is possible for spacetime geometry to permit retrograde time travel \cite{Lobo2010}. 
 
When discussing causality, we often describe the geometry in terms of the orientation of lightcones.  A lightcone emerging from a particular point can serve to demarcate the points which can and cannot be reached from that point along timelike trajectories. 

 In geometries which allow retrograde time travel, there are regions where our lightcones \emph{tip} in a way which lets our timelike worldlines spill backwards towards the ``past". In such regions of spacetime it is possible for  timelike curves to \emph{close} upon themselves --like some hackneyed snake eating its tail-- such curves are  described as being \emph{Closed Timelike Curves} (CTC).  The presence of CTC in a spacetime is used to demarcate those spacetimes where retrograde time travel is possible.

CTCs are most commonly generated as a consequence of angular momentum in a spacetime. The first geometry to be discovered to have CTC involves a  cosmological solution generated by rotating dust  \cite{Godel1949}. The Kerr spacetime and the  Tomimatsu-Sato spacetimes \cite{tomimatsu1973} all contain CTCs in their interior regions. CTCs can be generated by  infinitely long rotating cylinders of matter \cite{vanStockum1937}, (known as Tipler cylinders)\cite{Tipler1974}. Alternatively, Two cosmic strings passing one-another generate CTC \cite{Gott1991},\cite{Deser1992}. 

Beyond accidentally generating CTCs as a consequence of angular momentum, there are families of geometries which which have been deliberately designed to contain CTCs. The Ori \cite{Ori2005} and Ori-Soen \cite{Ori1993} spacetimes  both probe the physicality of generating a region with CTCs. If the two mouths of a traversable wormhole are properly accelerated with respect to one-another, the twin paradox can result in a CTC path between the mouths \cite{Echeverria1991,Friedman1990}.

\begin{figure}
\includegraphics[trim=0mm 0mm 0mm 0mm, clip, width=8cm]{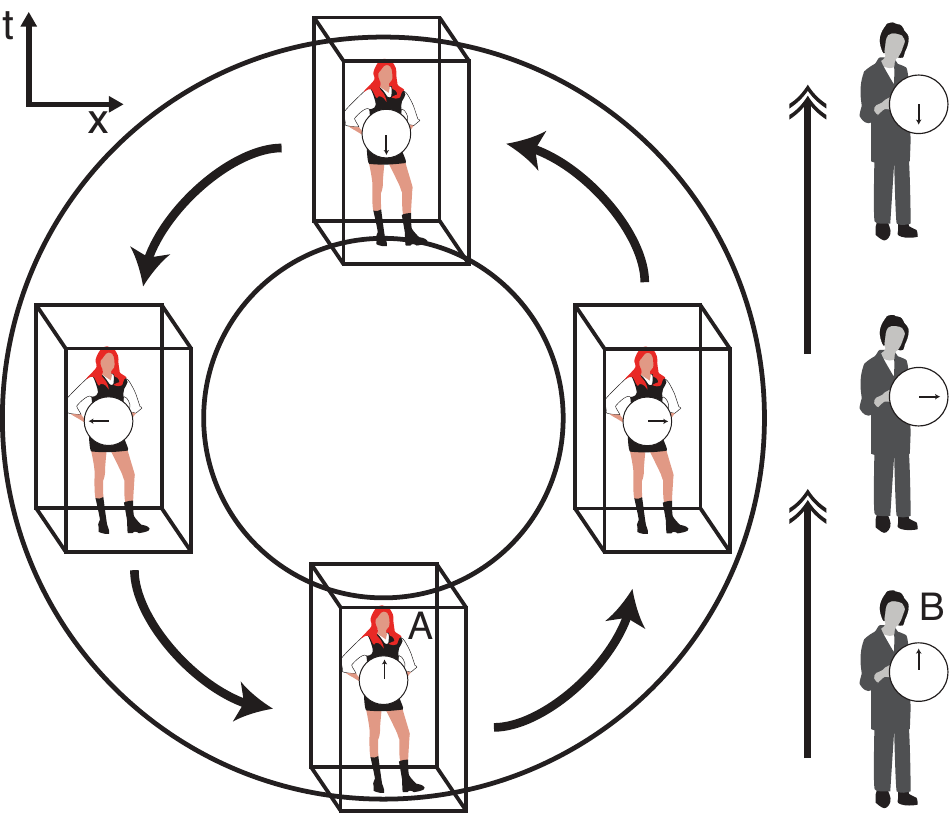}

\caption{A spacetime schematic of the trajectory of the TARDIS bubble. Arrows denote the local arrow of time. Companions Amy (A) and Barbara (B) experience different things inside and outside the TARDIS bubble. Within the TARDIS bubble, CTCs exist, and the local coordinate time follows the angle subtended by observer A as she travels in a circular trajectory. Outside the bubble, observer B sees time coordinate consistently increases in the vertical direction. Life inside the bubble is colourful and sexy and fun. Life outside the bubble is grey and drab, and dresses like a school teacher from the 1960's.  \label{clocks}}

\end{figure}

 \begin{figure}  \caption{Evolution of the boundaries of the bubble, as defined in section \ref{metron}, as seen by an external observer. At $T=-100$ the bubble will suddenly appear, and split in to two pieces which will move away from one another ($T=-75, \; T=-50$). At $T=0$, the two bubbles will come to rest, and then begin accelerating towards one-another ($T=+75, \; T=+50$). Whereupon, at $T=+100$ the two bubbles will merge and disappear.}
                             \label{fig:TEV}
                \centering
              \begin{subfigure}[b]{0.2\textwidth}
                             \includegraphics[trim=31mm 14mm 6mm 24mm, clip, width=4.3cm]{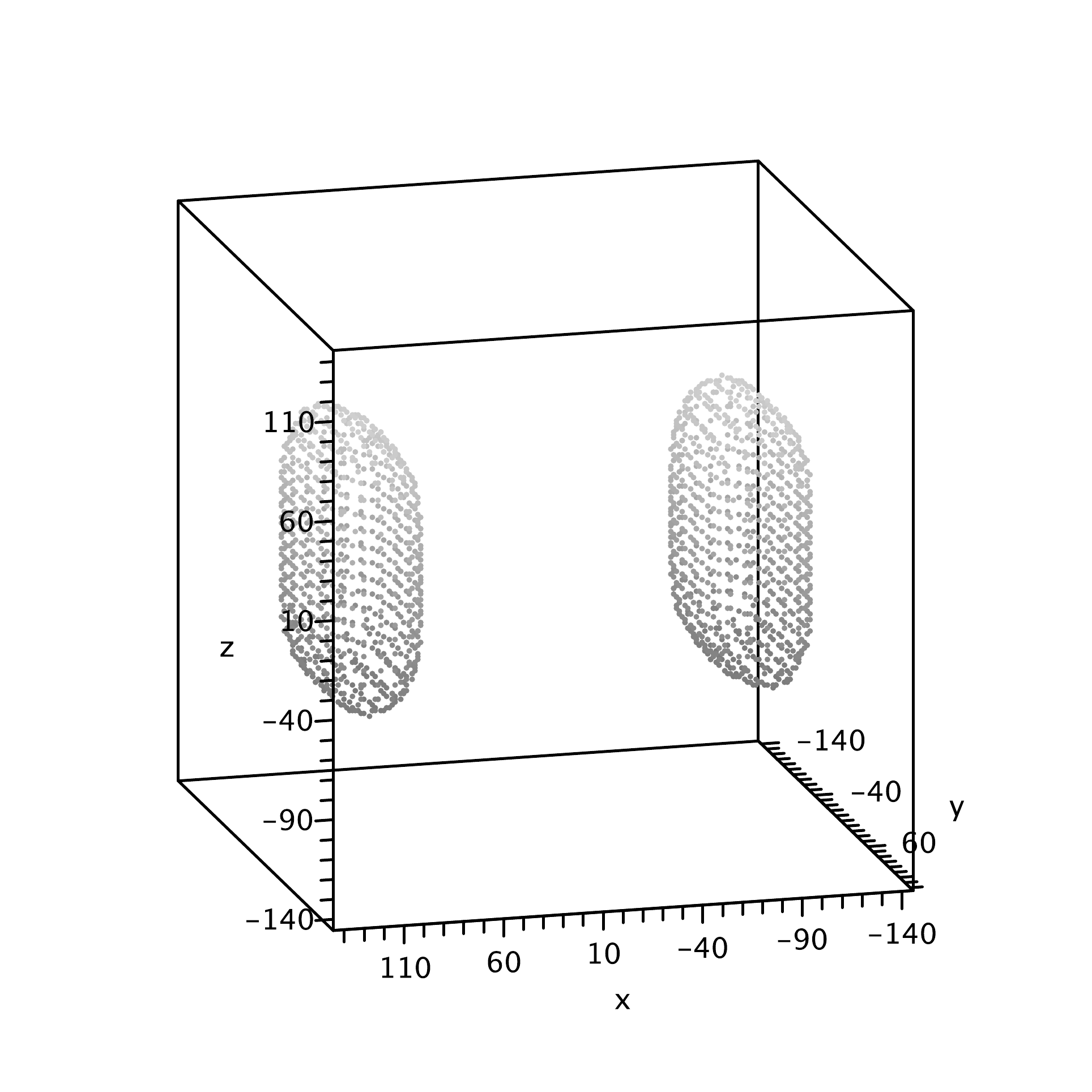}
                             \caption{TARDIS boundaries at T=0}
                             \label{fig:T0}
                   \end{subfigure}
                             \quad
                   \begin{subfigure}[b]{0.2\textwidth}
                             \includegraphics[trim=30mm 14mm 6mm 24mm, clip, width=4.3cm]{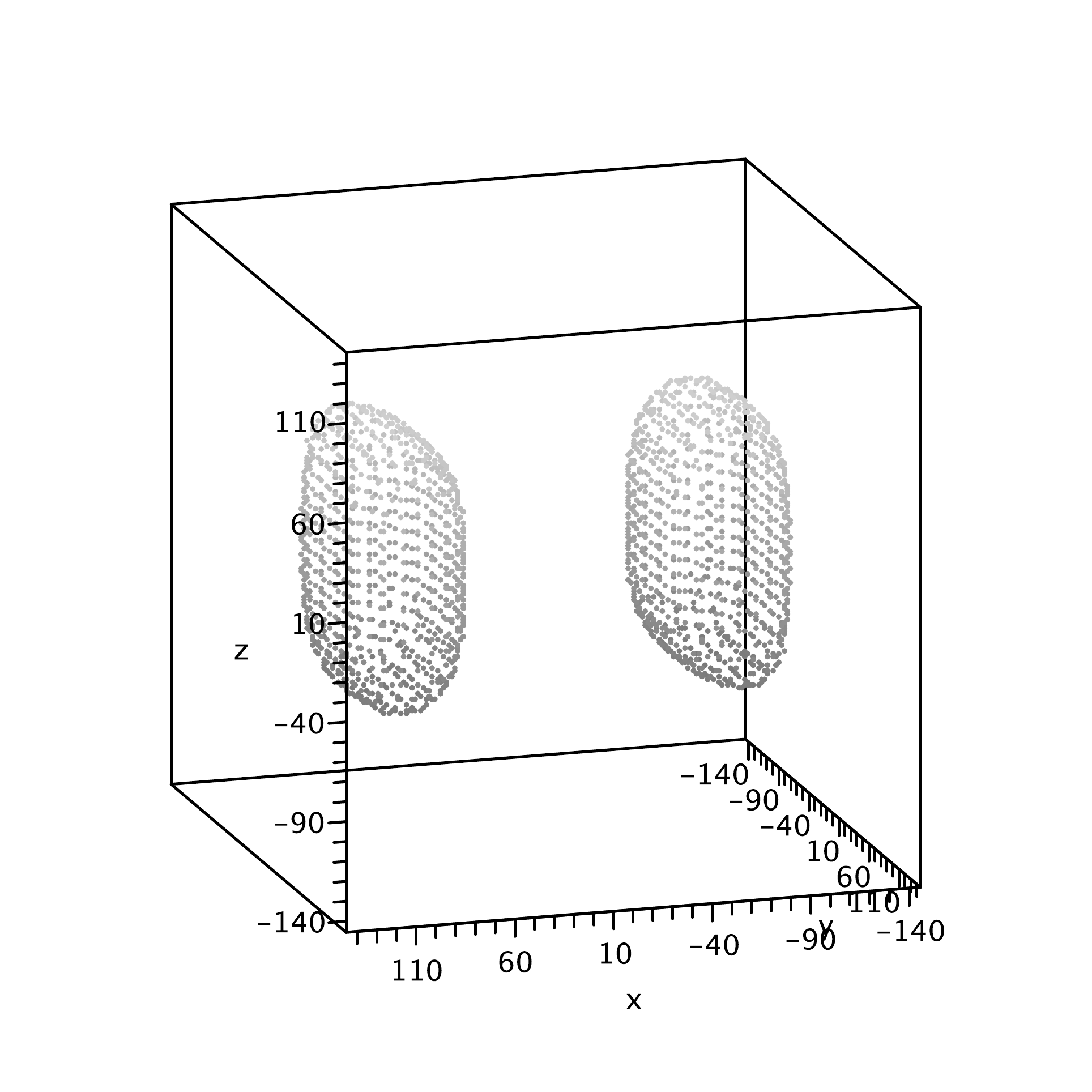}
                             \caption{TARDIS boundaries at T=$\pm 50$}
                             \label{fig:T0}
                   \end{subfigure}
                    \begin{subfigure}[b]{0.2\textwidth}
                             \includegraphics[trim=31mm 14mm 6mm 24mm, clip, width=4.3cm]{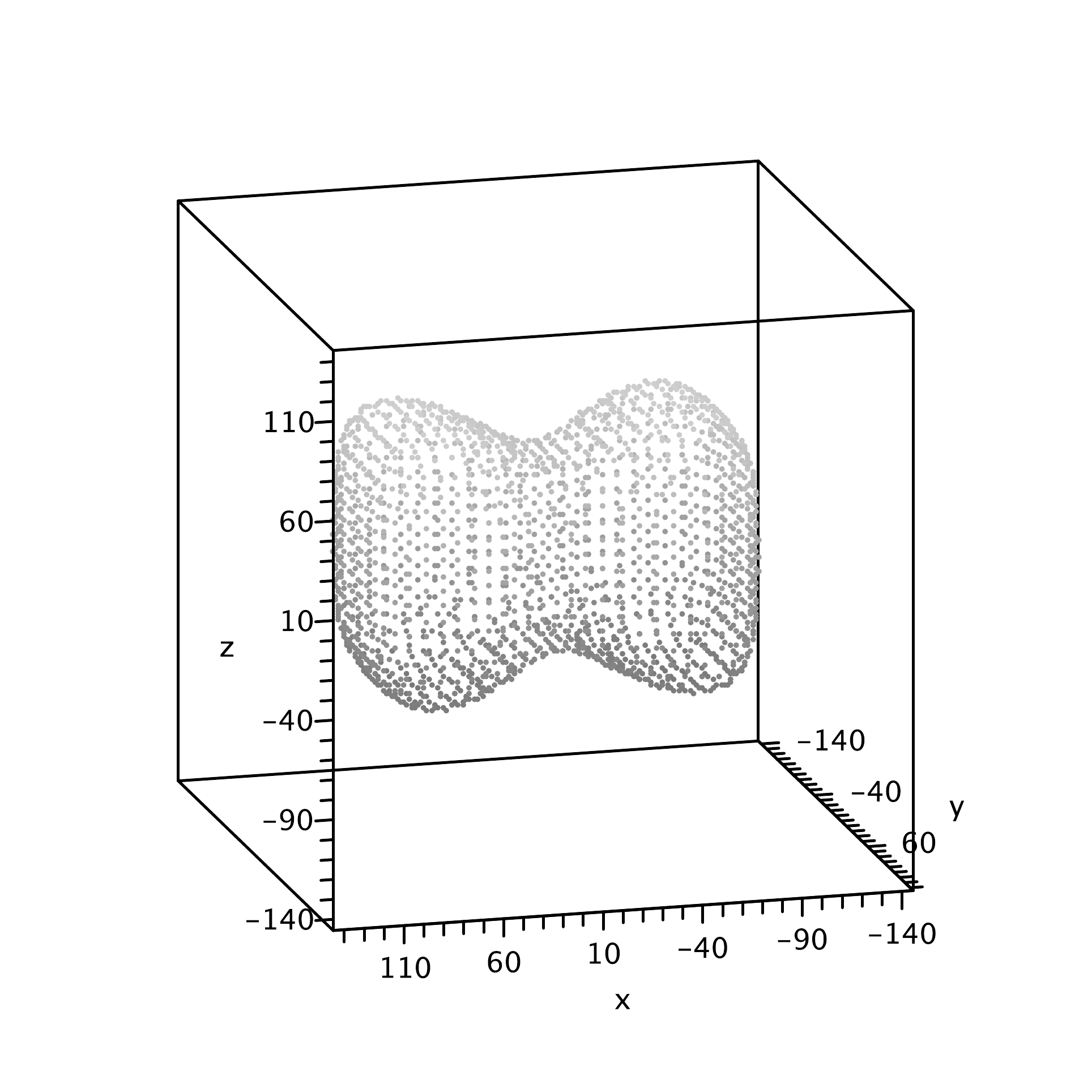}
                             \caption{TARDIS boundaries at T=$\pm 75$}
                             \label{fig:T0}
                   \end{subfigure}
                  \begin{subfigure}[b]{0.2\textwidth}
                             \includegraphics[trim=30mm 14mm 6mm 24mm, clip, width=4.3cm]{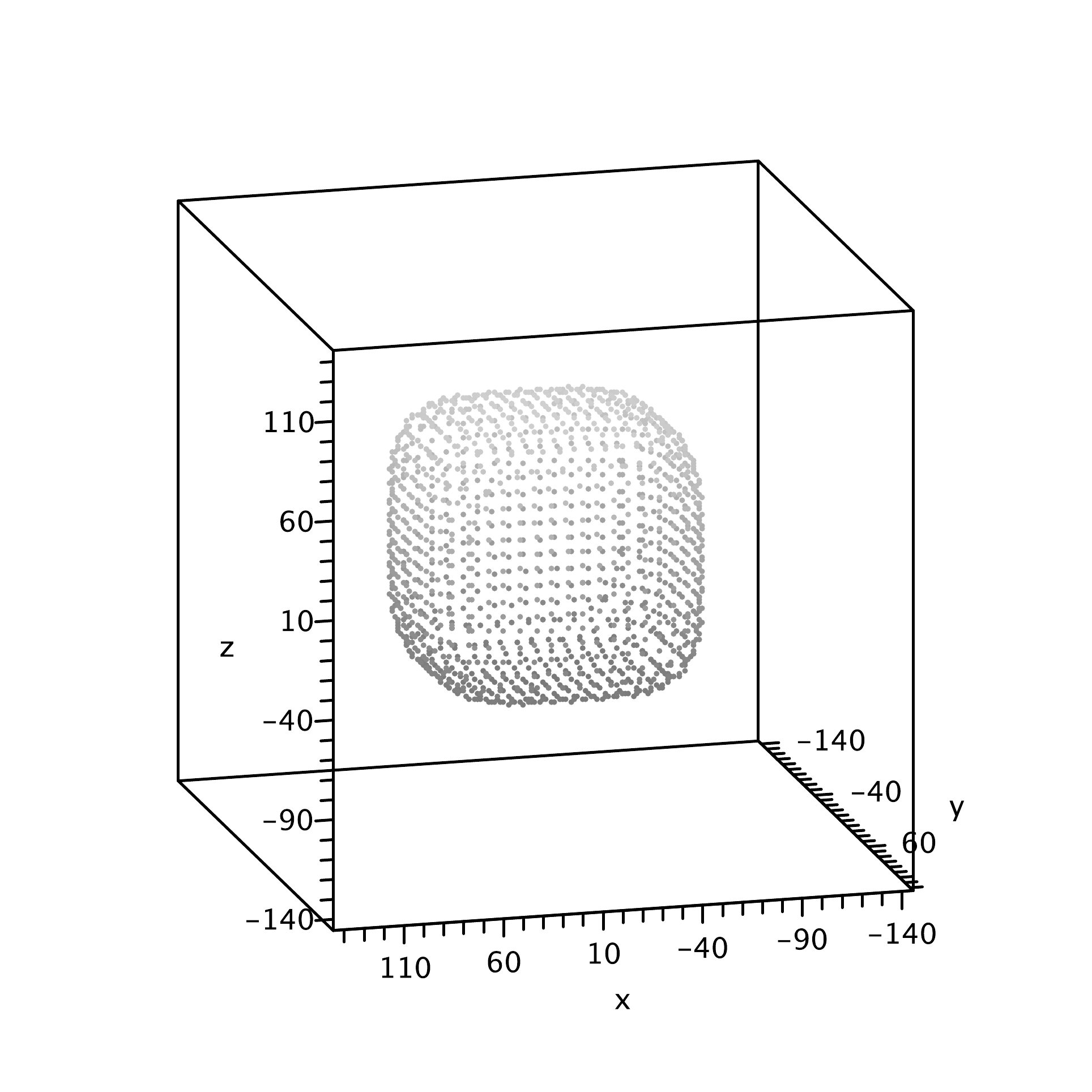}
                             \caption{TARDIS boundaries at T=$\pm 100$}
                             \label{fig:T0}
                   \end{subfigure}
 \end{figure}

Finally, spacetimes where timelike curves can travel along superluminal (as described by a distant observer) trajectories can be used to generate closed timelike curves.  The Alcubierre warp drive \cite{alcubierre94}  can be used to generate CTCs \cite{Everett1996},\cite{Gonzalez2000}; as can the Krasnikov tube \cite{Everett1997},\cite{Krasnikov1998}.

If there are so many ways to go about it, why have we never seen evidence of retrograde time travel in our universe? The answer depends on the particular CTC-containing geometry. Spacetimes permitting superluminal travel often require a violation of the classical energy conditions \cite{Lobo2010},\cite{Morris1988}, and are therefore not classically realizable.  The CTC region in the Kerr spacetime lay behind an event horizon, the Tomimatsu-Sato geometries are not recognized as the endstate of gravitational collapse, and the other  mentioned models require infinitely large distributions of matter. 

Alternatively, there is an argument, due to Hawking \cite{Hawking1991}, known as the \emph{chronology protection conjecture}. Hawking's arguments are centred around how most spacetimes which go from having no CTC at an early time, to possessing them at a later time,  also possess compactly generated Cauchy horizons. He argues that the energy density in a realistic spacetime will diverge immediately prior to the formation of CTCs along these surfaces (thus, distorting the geometry, and preventing their formation)\cite{Hawking1991}.

In spite of this menagerie of CTC geometries, a layperson might ask whether any of these solutions can be described as the oft-dreamt-of time machine? Certainly, retrograde time travel is possible in all of these geometries. Conversely, the Tipler cylinder \cite{Tipler1974} bares as much semblance to a box which travels backwards in time as an avalanche resembles an snow mobile. 

The purpose of this paper is to present a spacetime geometry which we might simply describe as a box which travels backwards and forwards along a loop in spacetime. Similar to the Alcubierre drive, it is a bubble of curved geometry  embedded in a flat, Minkowski vacuum. Unlike the Alcubierre drive,  the trajectory of our bubble is closed in spacetime. 

\subsection{The Traversable Achronal Retrograde Domain}

Our spacetime contains a bubble of  geometry which travels along a closed (circular) path through Minkowski spacetime. Accelerating observers within the bubble may travel along timelike curves which close within the bubble's worldtube. 

An observers travelling within the bubble will tell a dramatically different story from observers outside of it. To illustrate the general properties of this spacetime, let us imagine a pair of companions: one named Amy (A), who rides within the bubble; and the other named Barbara (B), who is left behind and remains outside of it. Let us also suppose that the walls of the bubble are transparent, and both observers are holding aloft large clocks (see Fig. \ref{clocks}).

How will Amy describe her bubble's trajectory? Counter-clockwise in spacetime! Initially, it begins moving at a subluminal speed to the left (evolving forwards in time relative to the Minkowski exterior).  The bubble then begins to accelerate until it is moving superluminally (to \emph{turn around} in time). The bubble then slows down to subluminal speeds, but now it is moving backwards in time relative to the Minkowski exterior. Finally, it begins to accelerate  towards the right ( towards its initial location) until it is again moving superluminally, whereupon it decelerates and its worldtube closes.  Amy will see the hands of her own clock move clockwise the entire time. When she looks out at Barbara's clock, she will see it alternate between moving clockwise  and moving counter-clockwise. 
 
Barbara, looking upon Amy, will see something rather peculiar. At some initial time ($T=-100$ in Fig.  \ref{fig:TEV}), two bubbles will suddenly appear, moving away from one-another. The two bubbles will slow down, stop (at $T=0$), and then begin moving back towards each other, until they merge (at $T=+100$) . Amy will appear in both of the bubbles:  the hands of the clock will turn clockwise in one bubble, and counter-clockwise in the other. 

Note that Barbara's interpretation of events is not unlike that used to describe the pair-creation and subsequent annihilation of a positron-electron pair: we often describe the positron as being an electron which is moving backwards in time. 

Due to the unique features of this bubble  geometry, we refer to it as a \emph{Traversable Achronal Retrograde Domain In Spacetime} (TARDIS), and we will proceed to describe its geometry below.

\section{TARDIS Geometry}
\subsection{Metric \label{metron}}

\begin{figure*}
\includegraphics[scale=0.8]{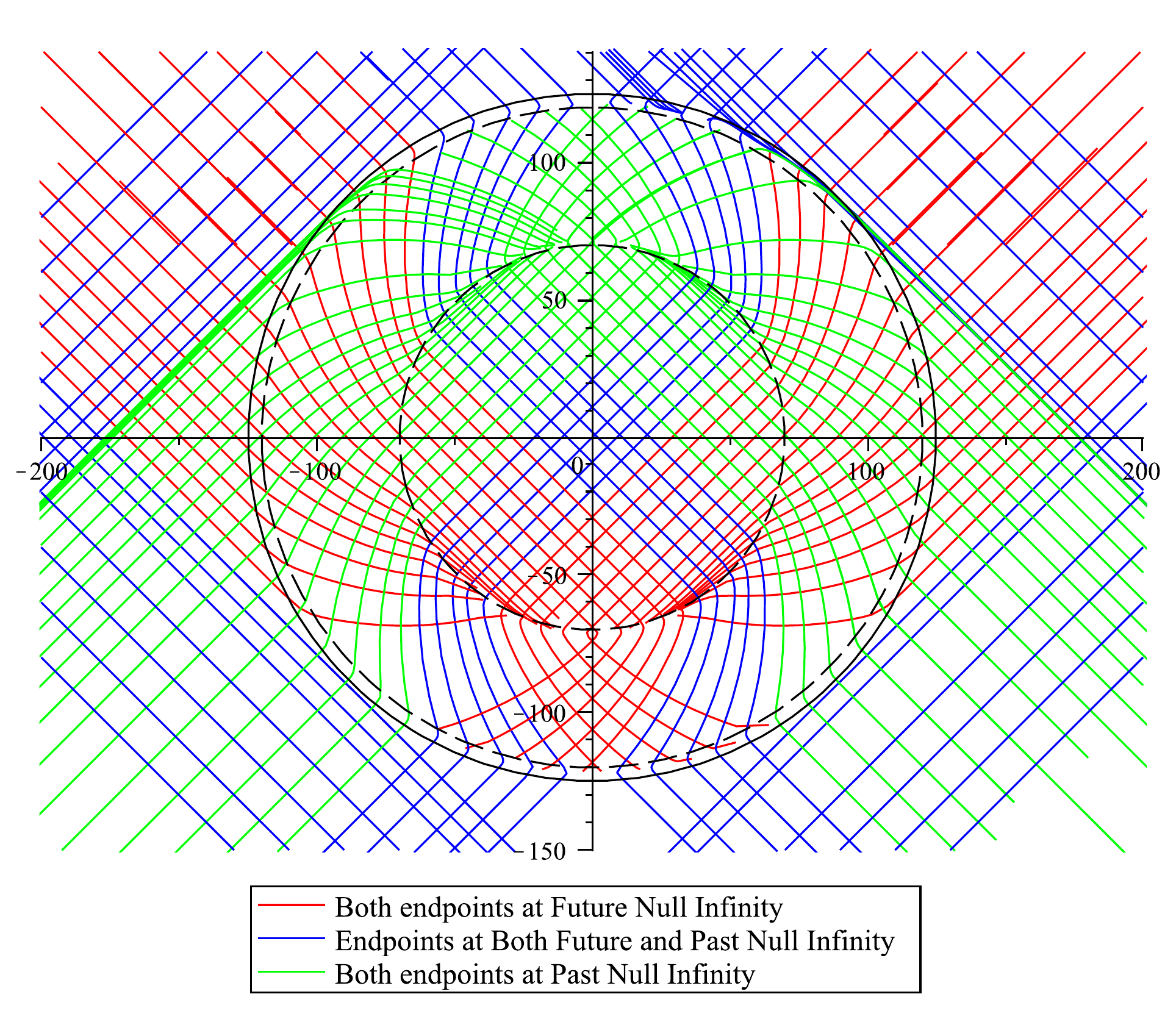}

\caption{ Null geodesics travelling through a cross section $y=0,\,z=0$. Black circles represent the edges of the bubble. Green curves get reflected by the bubble geometry and will not cross any Spacelike hypersurfaces which lie far to the future of the bubble. Red curves are likewise reflected, and will not cross any spacelike hypersurfaces which lie far the to past of the bubble. Blue curves will cross both future and past spacelike hypersurfaces. Note also how lightcones within the bubble \emph{tilt} to allow timelike observers to travel along closed, circular worldlines.  \label{hello}}

\end{figure*}

\begin{figure}
\includegraphics[trim=140mm 40mm 80mm 40mm, clip, width=8cm]{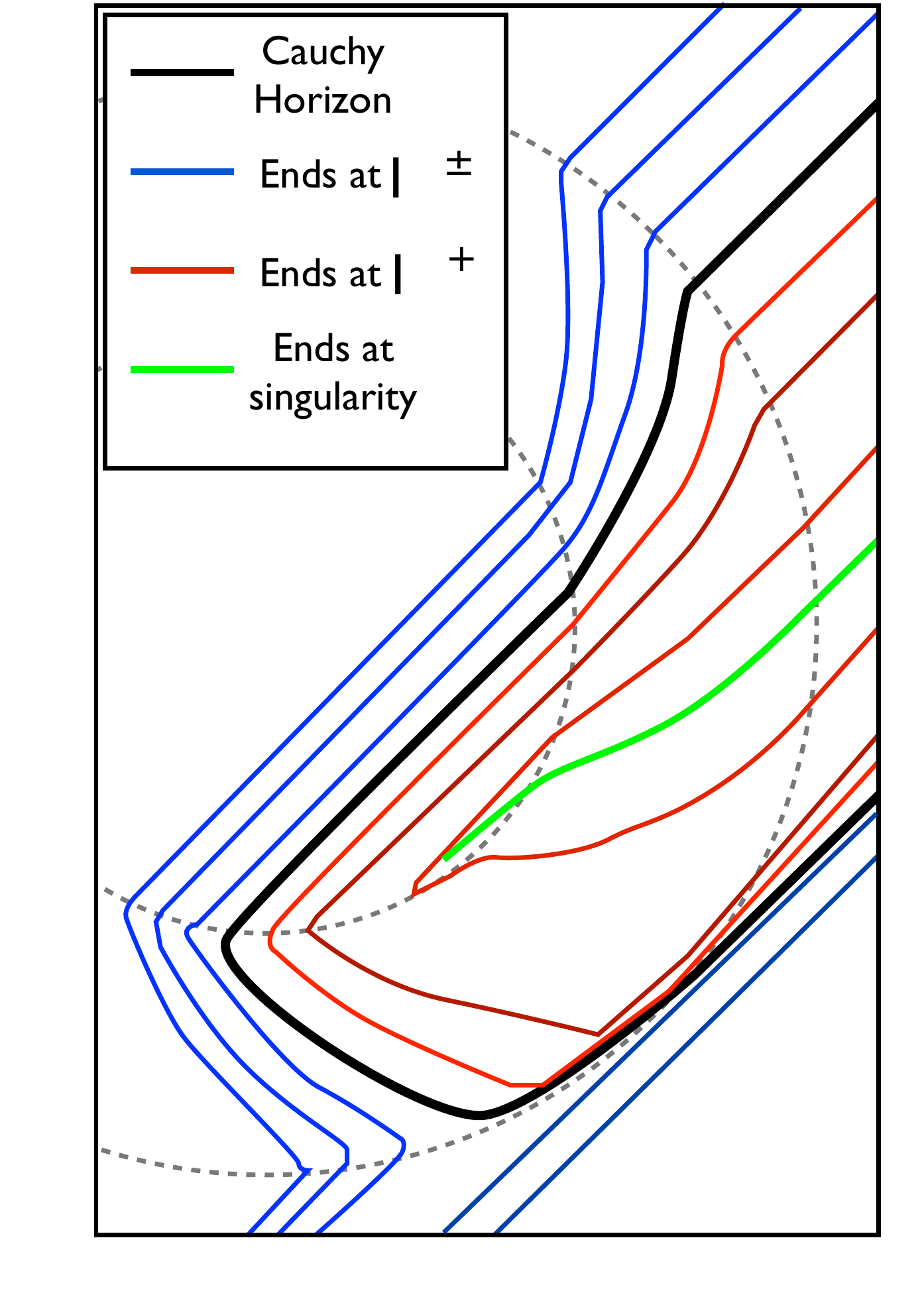}
\caption{ A schematic sketch of a set of parallel null curves being refracted by the edges of the moving bubble. Dashed lines represent the walls of the bubble. The Black curve is a generating curve of the Cauchy horizon. The $J^{\pm}$ are shorthand for future null and past null infinity, respectively. \label{cauchy}}

\end{figure}

The TARDIS geometry has the following metric:

 \begin{equation}\begin{split}
 ds^2= & \left[ 1 -     h(x,y,z,t)   \left( \frac{2t^2}{x^2 +t^2} \right)  \right] (-dt^2 +dx^2) \\
 & + h(x,y,z,t)  \left( \frac{4xt}{x^2 +t^2 } \right)dxdt +dy^2+dz^2 \label{eq:metric}
 \end{split} \end{equation}

Similar to the Alcubierre bubble, this metric relies  on a top-hat function $h(x,y,z,t)$ to demarcate the boundary between the inside of the bubble $h(x,y,z,t)=1$ and the exterior spacetime $h(x,y,z,t)=0$.

Clearly, on outside of the bubble (where $h(x,y,z,t)=0$), this metric is Minkowski spacetime. 

Let us consider the geometry which an observer inside of the bubble would see. Within this region: $h(x,y,z,t)=1$, and the metric can be re-written:
\begin{equation}
ds^2 =\left(  \frac{x^2 -t^2}{x^2 +t^2 } \right) (-dt^2 +dx^2)+ \left( \frac{4xt}{x^2 +t^2 } \right) dx dt + dy^2 +dz^2  \label{metricAlter}
\end{equation}
which, under the coordinate transformation:
\[ t= \xi \sin( \lambda ) \;, \; x=\xi  \cos( \lambda)
\] becomes Rindler spacetime.

\begin{equation}
ds^2=-\xi ^2 d\lambda^2+d\xi ^2 +dy^2 +dz^2  \label{rindler} 
\end{equation}

The Rindler geometry within this context has a modified topology, amounting to identifying the surfaces $\lambda=0$ with $\lambda =2\pi$. Thus, trajectories described with constant $\xi=\hat{A}, \, y=\hat{Y}, \, z=\hat{Z}$ coordinates will be closed timelike curves.
Note that while such curves are not geodesic, the acceleration required to stay on this trajectory can be modest. If $L^{a}=[\frac{1}{\xi},0,0,0]$ denotes a vector tangent to one of these CTC, and $K^{a}=[0,1,0,0]$ is an orthogonal spacelike vector:
\[
K^{a} L^{b} \nabla_{b} L_{a}=\frac{1}{\hat{A}} \; .\]
An observer moving along one of these curves will feel an acceleration equal to $1/\hat{A}$. Thus, the wider the ``radius" of the bubble's trajectory in spacetime, the weaker the acceleration required to travel along CTCs within it.

We shall define the edge of the TARDIS bubble to have  boxy spheroidal shape (with radial parameter $R$) moving along a circle of  ``radius" $A$ in the $x-t$ plane  (see Fig.  \ref{fig:TEV}):
\begin{equation}
h(x,y,z,t)=H \left( R^4 -z^4 -y^4 - \left[ x^2 +t^2 -A^2  \right]^{2} \right) \label{bubblo} 
 \end{equation}
where $H(x)$ denotes the Heaviside function.

We shall approximate the Heaviside function using a continuous Tanh function:
 \begin{equation}
 H(x)=\frac{1}{2}+\frac{\tanh(\alpha x)}{2} \label{lightside} \; .
 \end{equation}
 The $\alpha$ parameter can be set to define the suddenness of the transition between the inside and the outside of the bubble.
 
 All of the numerical models plotted in this paper will specify $A=100$, $R=70$ and $\alpha =\frac{1}{6000000}$.

  \begin{figure}   \caption{Nonzero terms of the stress energy tensor along the slice $y=0,\,z=0$.}
                             \label{fig:y0z0}
                \centering
              \begin{subfigure}[b]{0.4\textwidth}
                             \includegraphics[trim=13mm 5mm 6mm 4mm, clip, width=8cm]{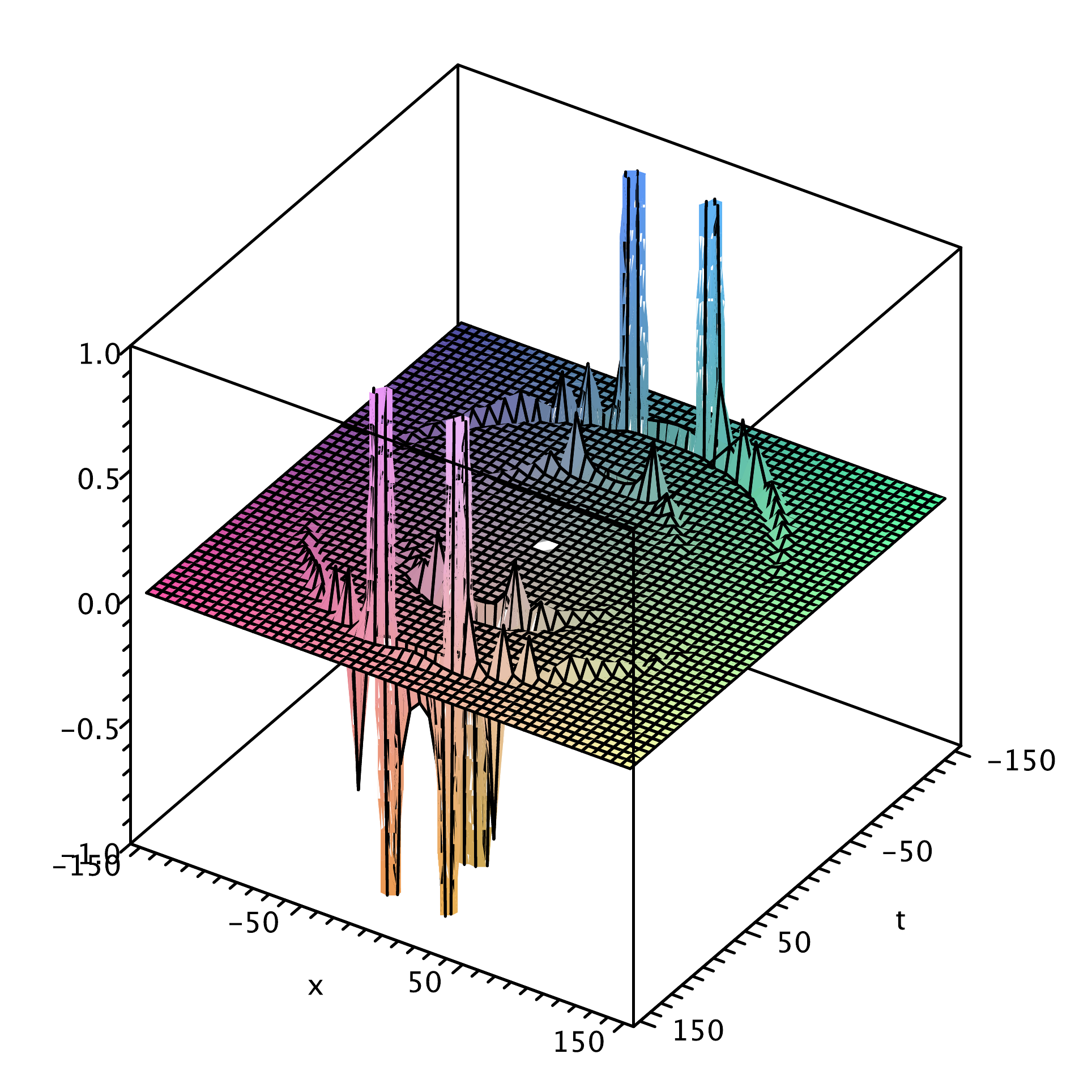}
                             \caption{$G_{zz}$ along the slice $y=0,\,z=0$}
                             \label{fig:p1}
                   \end{subfigure}
              \begin{subfigure}[b]{0.4\textwidth}
                             \includegraphics[trim=13mm 5mm 6mm 4mm, clip, width=8cm]{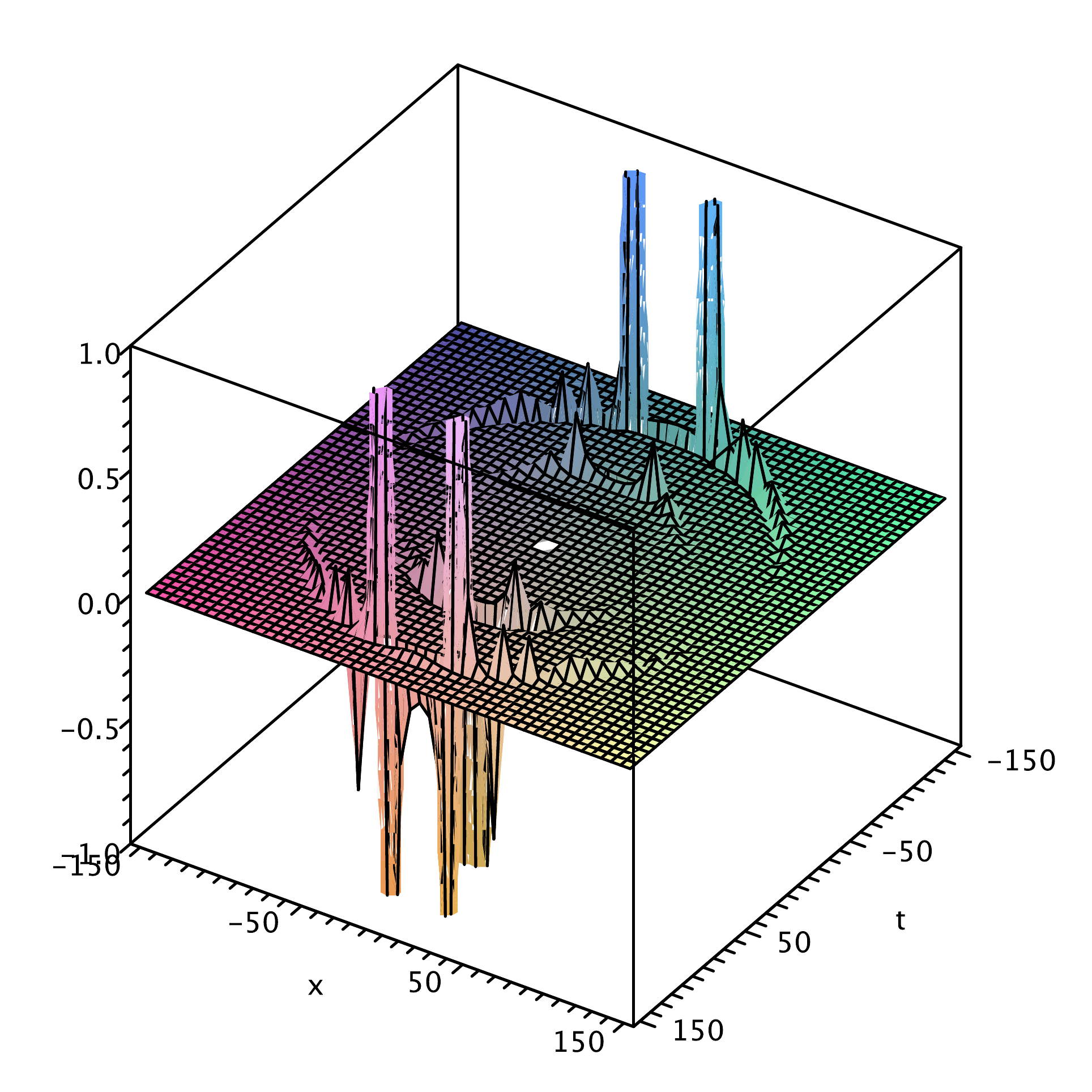}
                             \caption{$G_{yy}$ along the slice $y=0,\,z=0$}
                             \label{fig:p2}
                   \end{subfigure}
                   
 \end{figure}
 
 \begin{figure} \caption{Nonzero terms of the stress energy tensor along the slice $y=0,\,t=0$}
                             \label{fig:y0t0}
                \centering
              \begin{subfigure}[b]{0.29\textwidth}
                             \includegraphics[trim=14mm 5mm 6mm 4mm, clip, width=7cm]{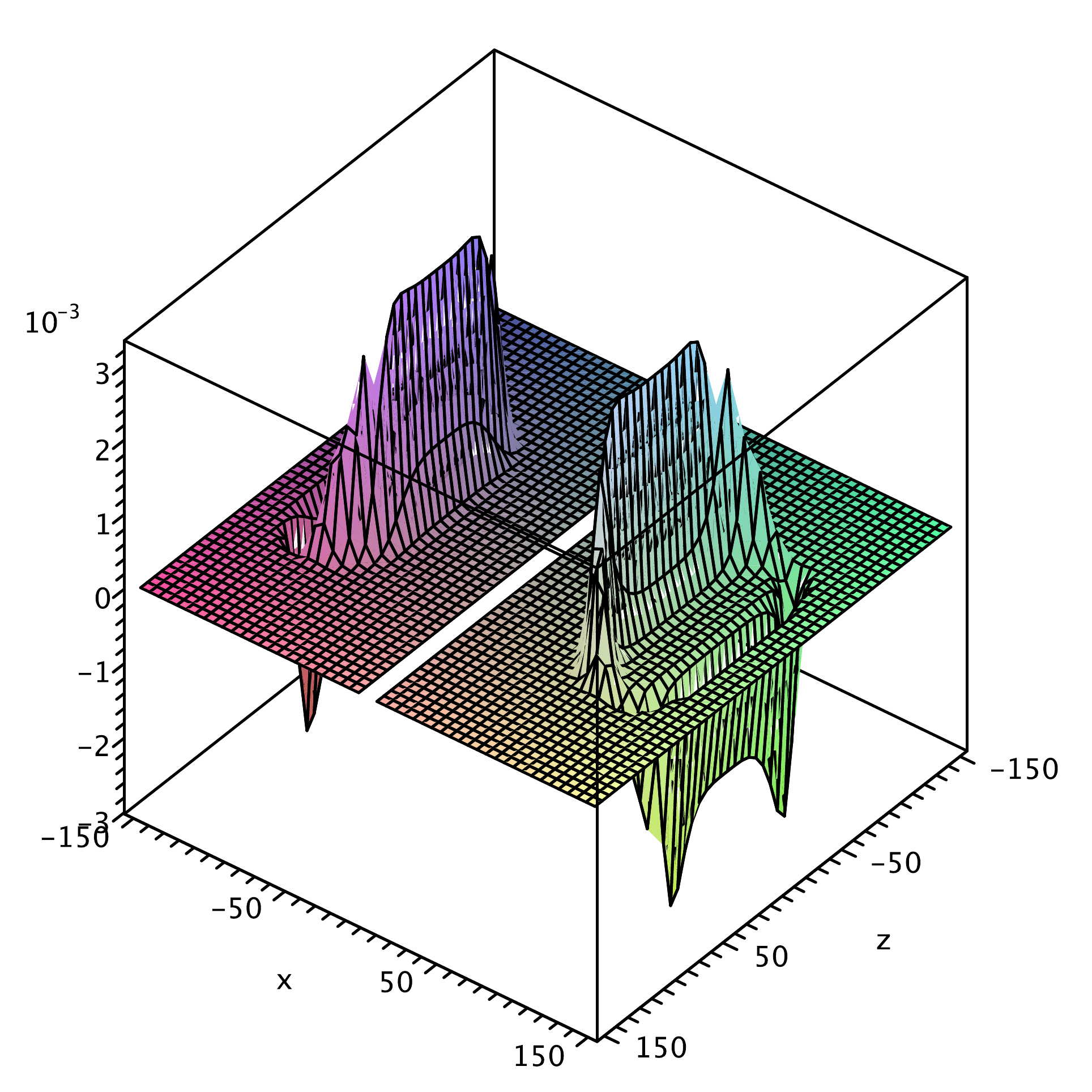}
                             \caption{$G_{zz}$ along the slice $y=0,\, t=0$}
                             \label{fig:p3}
                   \end{subfigure}
              \begin{subfigure}[b]{0.29\textwidth}
                             \includegraphics[trim=14mm 5mm 6mm 4mm, clip, width=7cm]{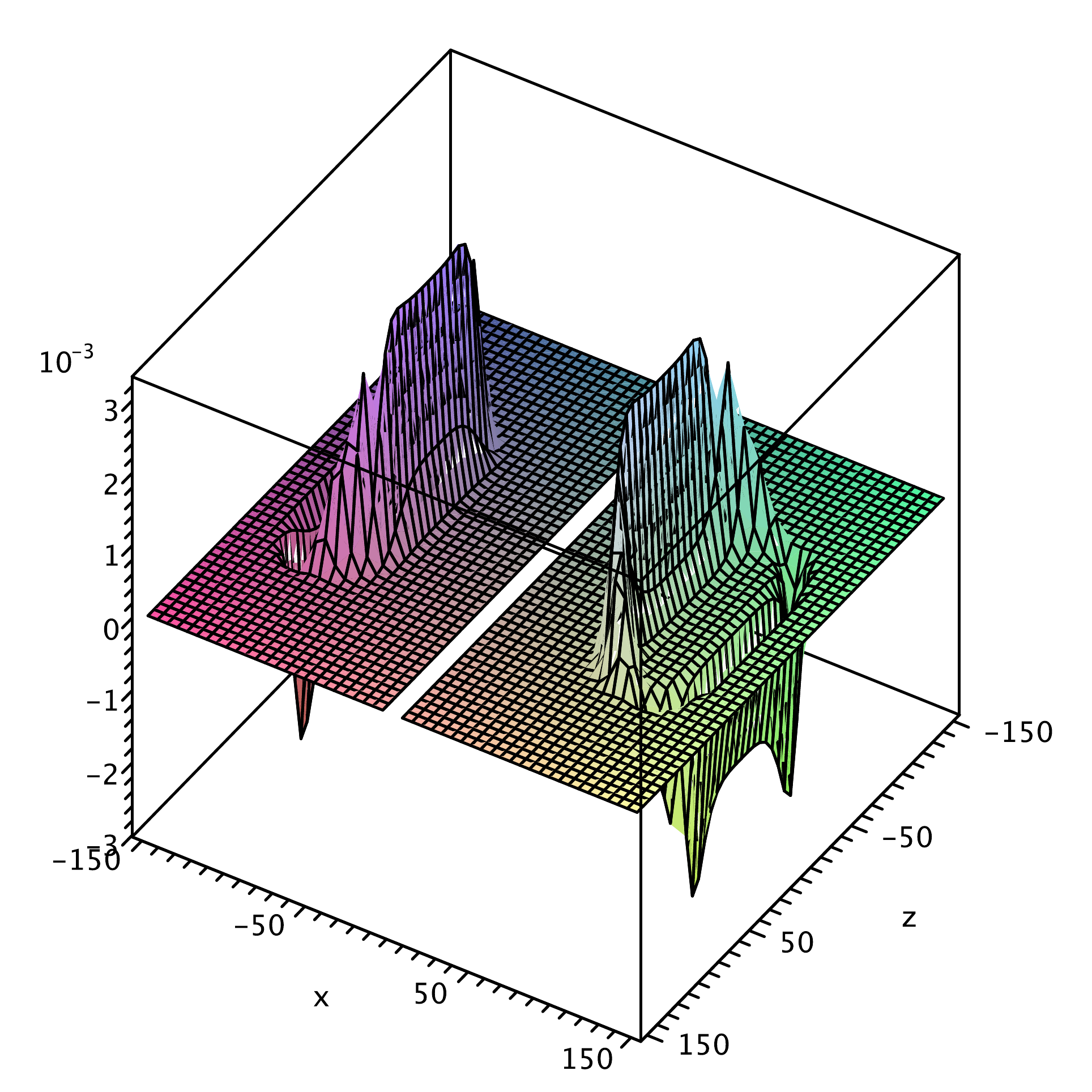}
                             \caption{$G_{yy}$ along the slice $y=0,\, t=0$}
                             \label{fig:p4}
                   \end{subfigure}
              \begin{subfigure}[b]{0.29\textwidth}
                             \includegraphics[trim=14mm 5mm 6mm 4mm, clip, width=7cm]{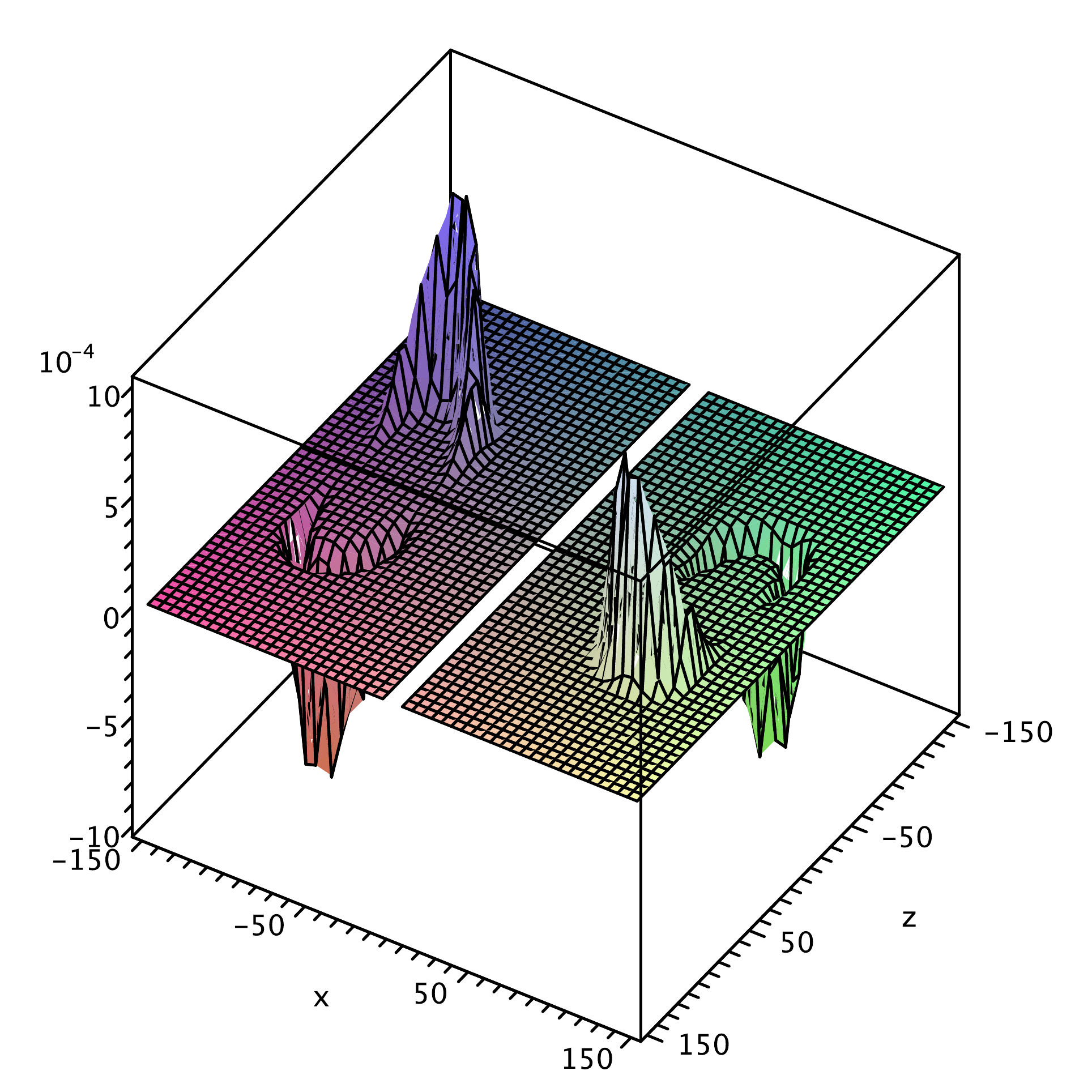}
                             \caption{$G_{xz}$ along the slice $y=0,\, t=0$}
                             \label{fig:p5}
                   \end{subfigure}
                   
 \end{figure}

 \subsection{Causal Structure \label{causalstru}}

  In order to probe the causal structure of a TARDIS bubble geometry, in Fig.  \ref{hello} we have ray-traced null geodesics across the slice $y=0$, $z=0$. The intersection of any two null geodesics illustrate how the lightcones are tilted. Along the x-axis ($t=0$), all lightcones are pointed up-down. Note that as we move along the inside of the bubble, lightcones tip in an appropriate manner to allow circular CTC trajectories. 
 
Interestingly, an observer within the bubble will be able to see past (or future) versions of themselves travelling within the bubble. 
 
Null geodesics, when they encounter the bubble walls, may diffract strongly. As a result, some of the null curves will be effectively ``turned around in time" (see Fig. \ref{cauchy}).  Thus, while some of the geodesics in Fig. \ref{hello} have end points on future  and  past null infinity (coloured blue), others have endpoints \emph{only} at future null infinity (red), or \emph{only} at past null infinity (green). This illustrates the fact that no spacetime containing a TARDIS bubble can be globally hyperbolic, and that the bubble must be the source of a Cauchy horizon.
 
 Consider the schematic in Fig. \ref{cauchy}, which illustrates these various behaviours. The red curves are those which originate from past null infinity and either miss the bubble or diffract through it, and then end at future null infinity. The red curves get strongly diffracted by the bubble, and both endpoints lay at future null infinity. Of these curves, the geodesics which bound this set (in black) are among those geodesics which \emph{generate} the Cauchy horizon. They are the earliest null geodesics to emerge from the bubble. We note that these geodesics are not compactly generated, and we believe that Hawking's chronology protection conjecture will not apply to this spacetime \cite{Hawking1991,Kim1991} . 
 
 Note finally that, at the centre of the family of the null curves which have been folded over in spacetime (Fig.  \ref{cauchy}), there must be one null curve which will be folded onto itself, and therefore cannot be extended to arbitrary parametric length. It is not clear whether this endpoint should be interpreted as a singularity. We have not discovered any divergent behaviour in the geometry in the vicinity of this point.

 \subsection{Stress Energy Tensor}

 To give a sense of the matter required to transition between the interior and the exterior vacuums of the bubble, we have plotted the various non-zero components of the stress energy tensor taken along cross sections of spacetime geometry. Consistently, the classical energy conditions are violated. 
 
 If we set $y=0, \; z=0$ we see a cross-section of the bubble as it travels along its circular course through spacetime. The only non-zero elements of the stress energy tensor along this slice are the non-zero pressures in the y and z directions, as seen in Fig.  \ref{fig:p1} and \ref{fig:p2} respectively. 
 
 If we set $y=0, \; t=0$ we see a cross-section of the bubble(s) at rest with respect to the exterior coordinate system.  There are non-zero pressure and also shear terms, as seen in Fig.  \ref{fig:p3},  \ref{fig:p4}, and \ref{fig:p5}.  
 
\begin{figure}  \caption{Nonzero terms of the stress energy tensor along the slice $x=100,\,t=0$.}
                             \label{fig:x100t0}
                \centering
              \begin{subfigure}[b]{0.4\textwidth}
                             \includegraphics[trim=14mm 5mm 6mm 4mm, clip, width=8cm]{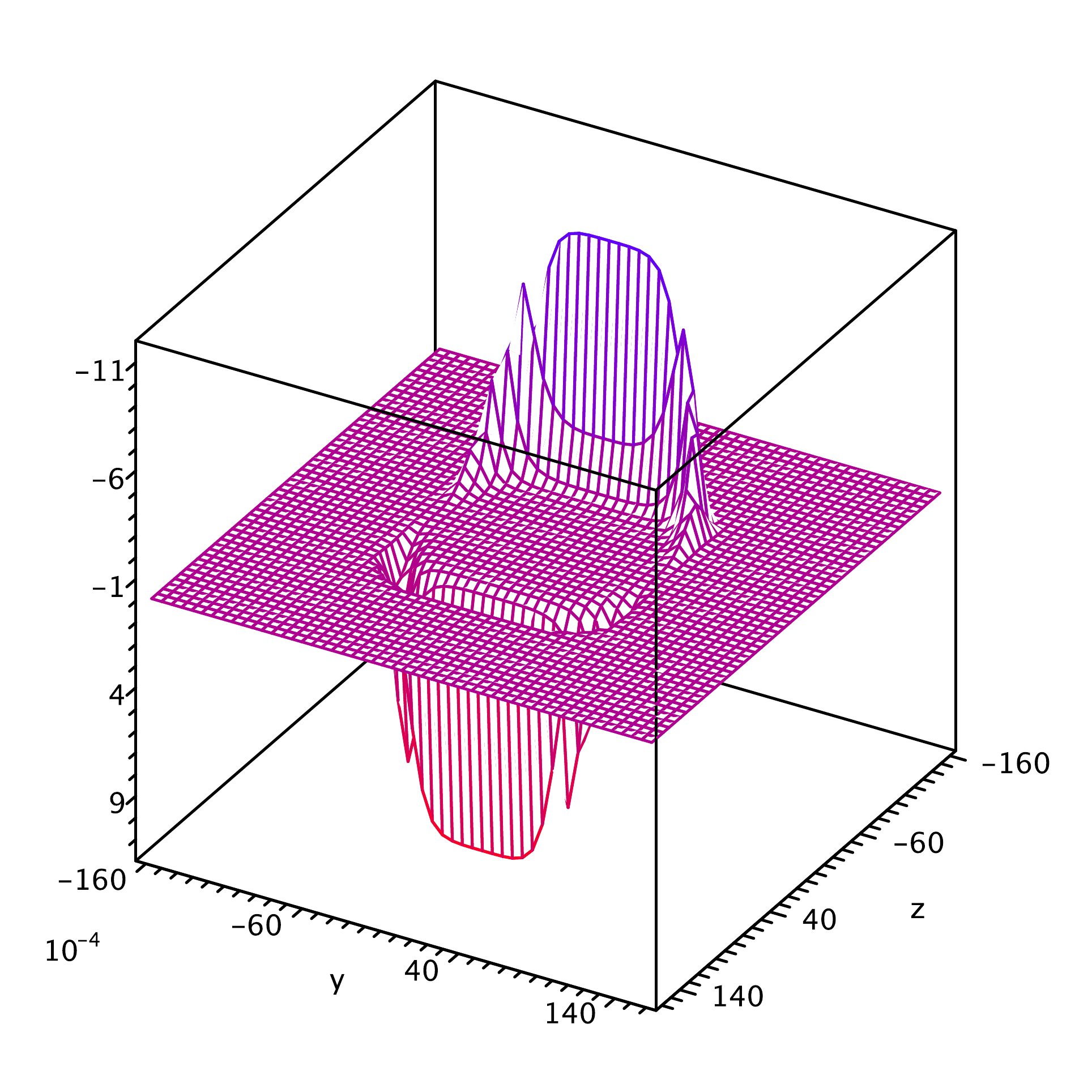}
                             \caption{$G_{xz}$ along the slice $t=0, \; x=100$}
                             \label{fig:p6}
                   \end{subfigure}
              \begin{subfigure}[b]{0.4\textwidth}
                             \includegraphics[trim=14mm 5mm 6mm 4mm, clip, width=8cm]{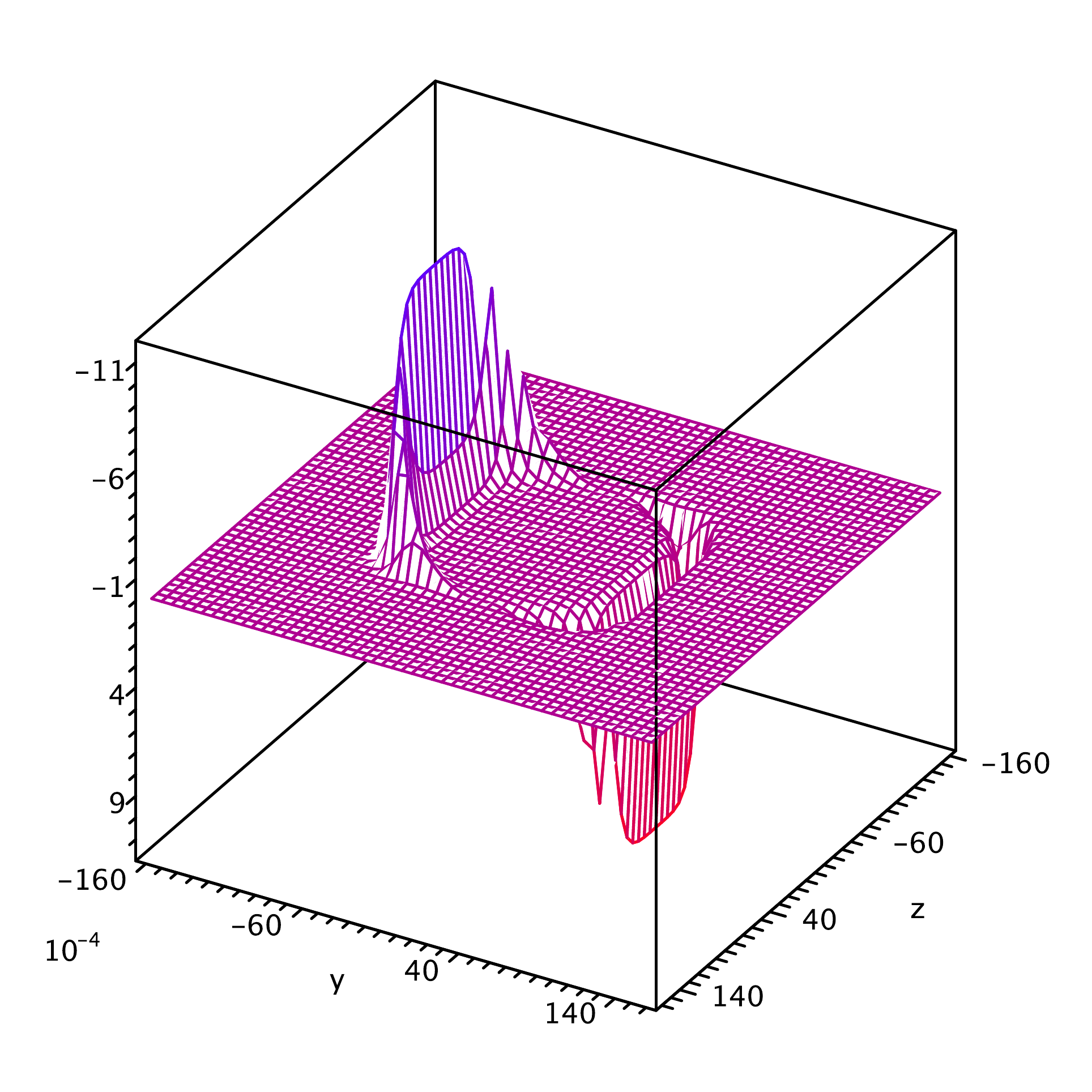}
                             \caption{$G_{xy}$ along the slice $t=0, \; x=100$}
                             \label{fig:p7}
                   \end{subfigure}
                   
 \end{figure}

 If we set $t=0, \; x=100$ we see a cross-section of the bubble over the y-z plane when the bubble is at rest relative to the exterior coordinate system.  In this case, the only non-zero elements of the stress energy tensor along this slice are shear terms, as seen in Fig.  \ref{fig:p6} and \ref{fig:p7}.

\begin{figure*}   \caption{Nonzero elements of the stress energy tensor along the slice $y=0,\,z=30$.}
                            \label{fig:p7x}
                \centering
               \begin{subfigure}[b]{0.3\textwidth}
                             \includegraphics[trim=14mm 5mm 6mm 4mm, clip, width=6cm]{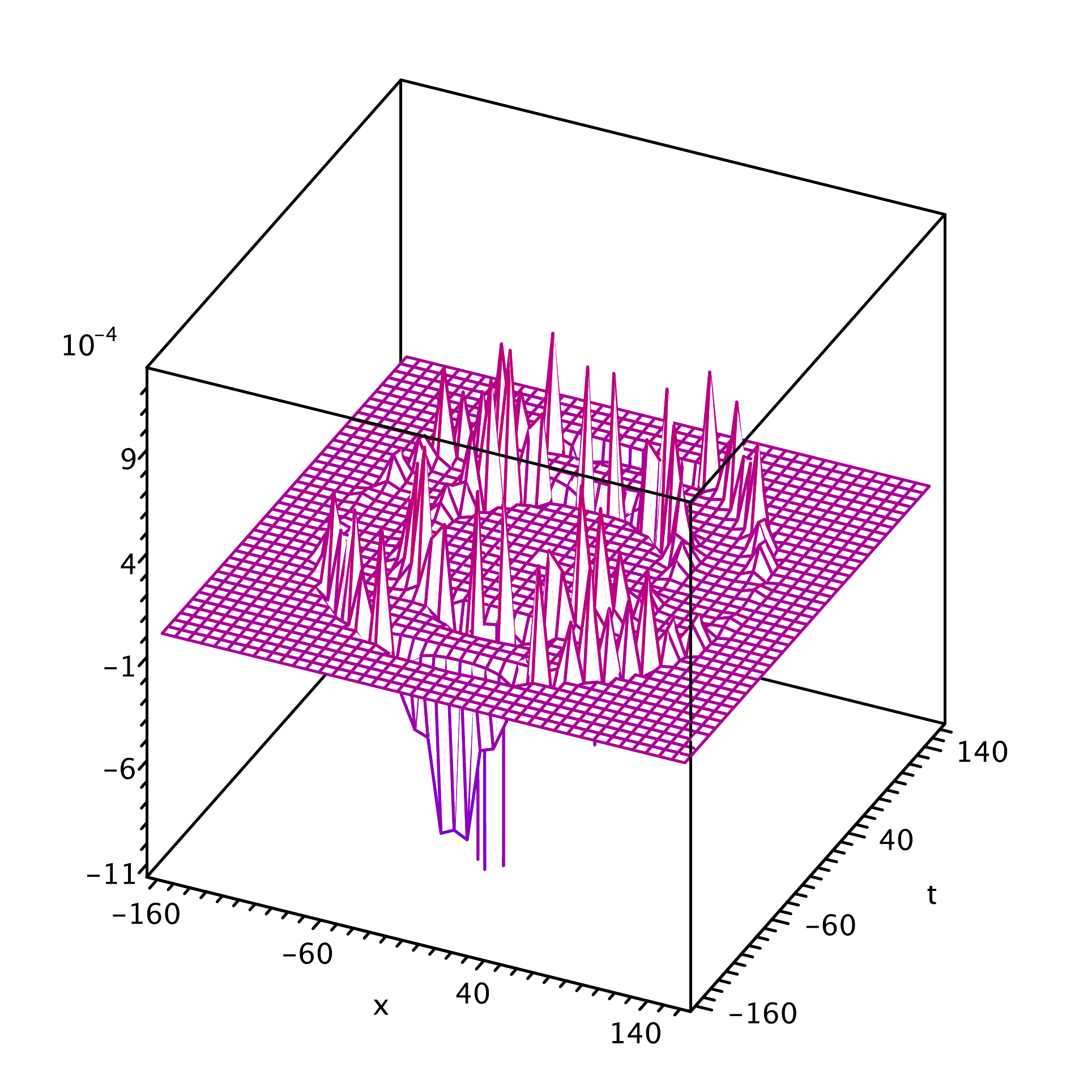}
                             \caption{$G_{tt}$ along the slice $z=30, \; y=0$}
                   \end{subfigure}
           \begin{subfigure}[b]{0.3\textwidth}
                             \includegraphics[trim=14mm 5mm 6mm 4mm, clip, width=6cm]{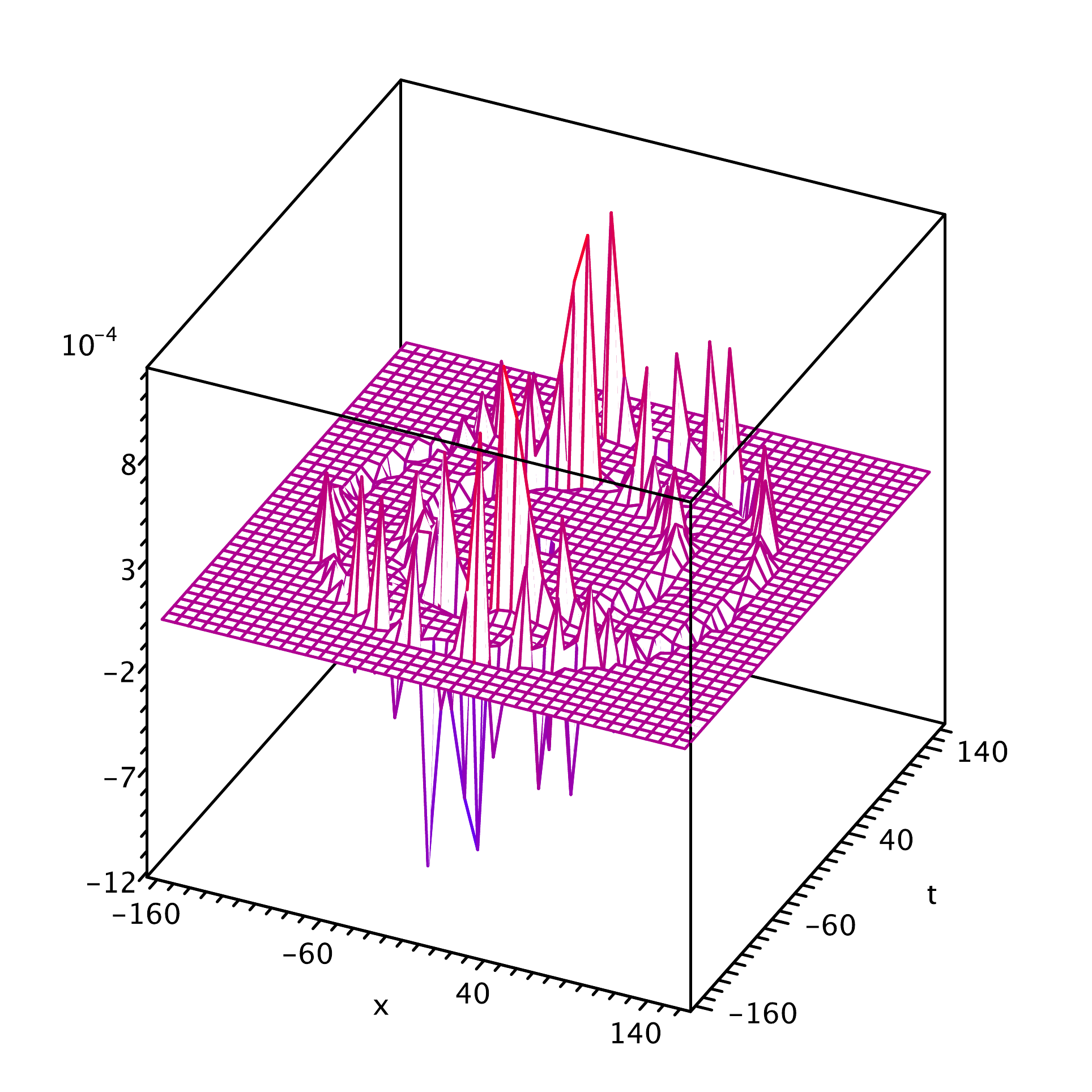}
                             \caption{$G_{ty}$ along the slice $z=30, \; y=0$}
                   \end{subfigure}
                   \begin{subfigure}[b]{0.3\textwidth}
                             \includegraphics[trim=14mm 5mm 6mm 4mm, clip, width=6cm]{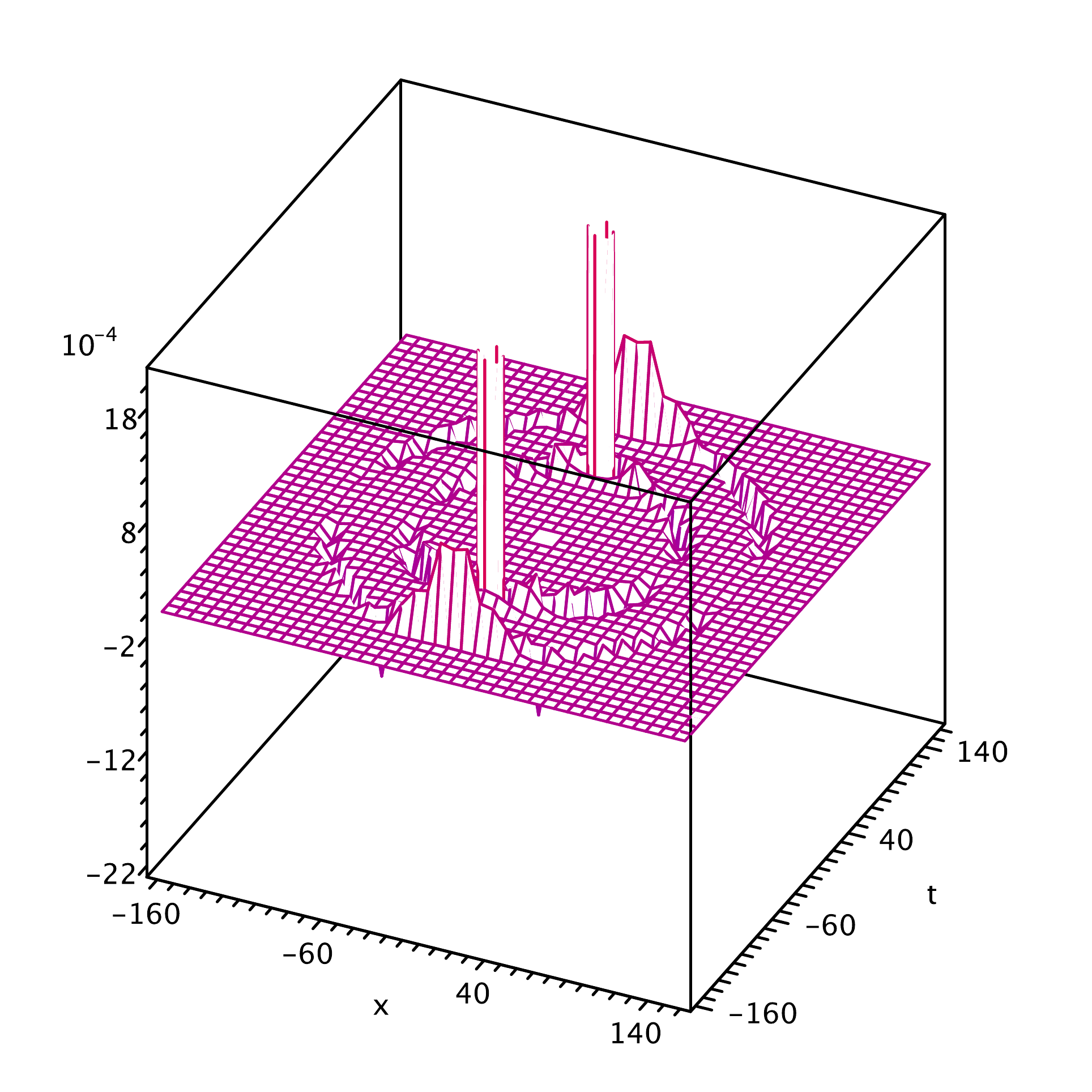}
                             \caption{$G_{xx}$ along the slice $z=30, \; y=0$}
                   \end{subfigure}
             \begin{subfigure}[b]{0.3\textwidth}
                             \includegraphics[trim=14mm 5mm 6mm 4mm, clip, width=6cm]{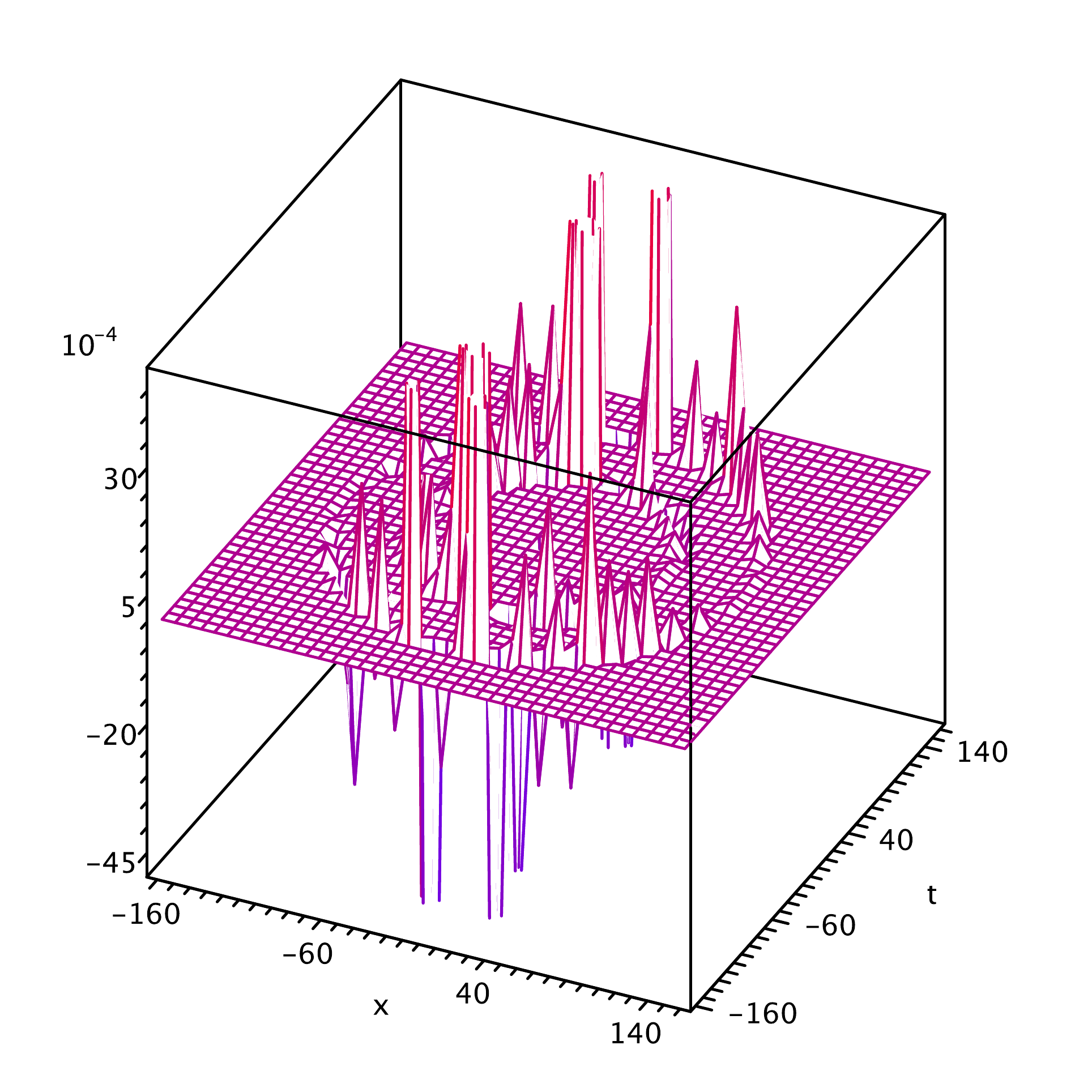}
                             \caption{$G_{xz}$ along the slice $z=30, \; y=0$}
                   \end{subfigure}
           \begin{subfigure}[b]{0.3\textwidth}
                             \includegraphics[trim=14mm 5mm 6mm 4mm, clip, width=6cm]{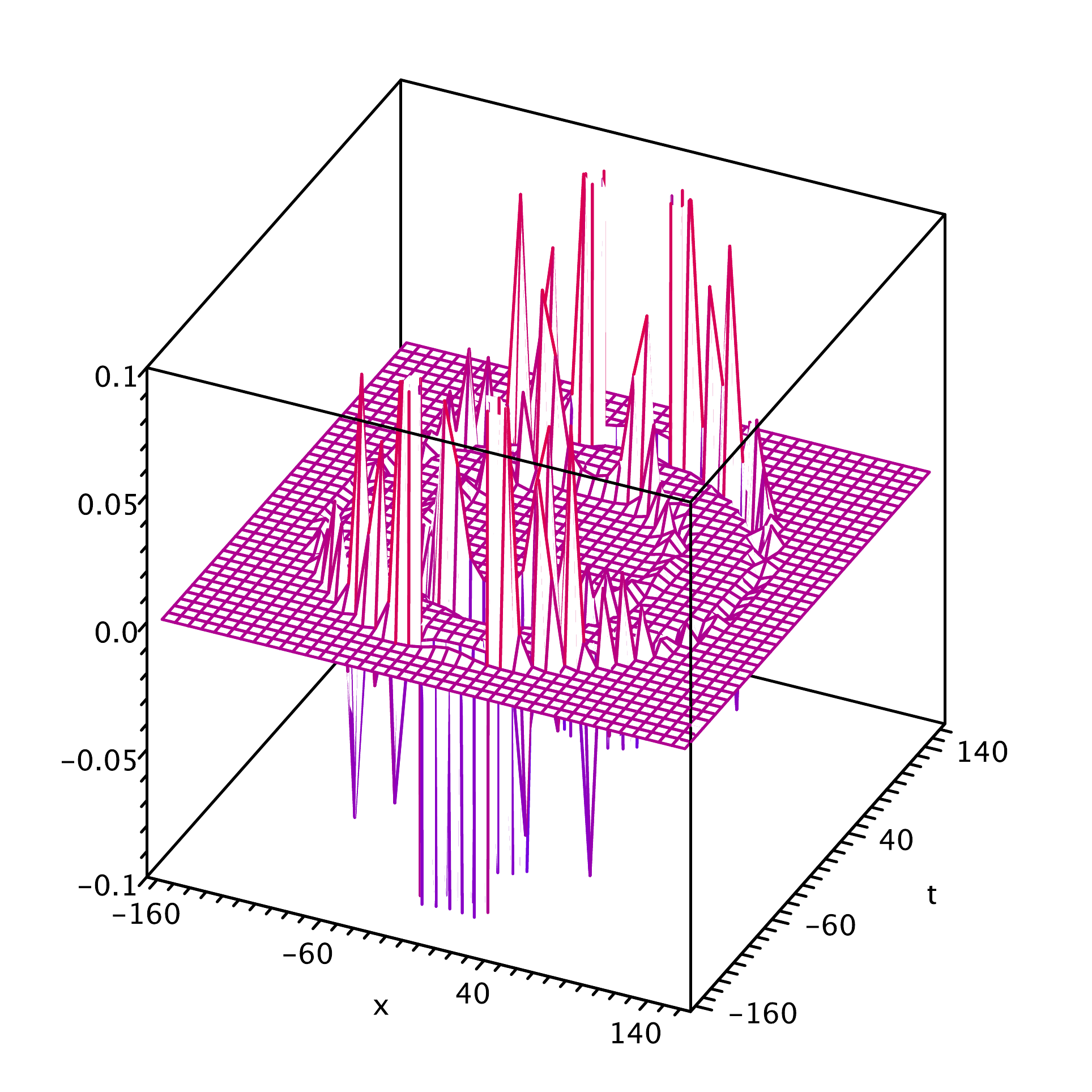}
                             \caption{$G_{yy}$ along the slice $z=30, \; y=0$}
                   \end{subfigure}
              \begin{subfigure}[b]{0.3\textwidth}
                             \includegraphics[trim=14mm 5mm 6mm 4mm, clip, width=6cm]{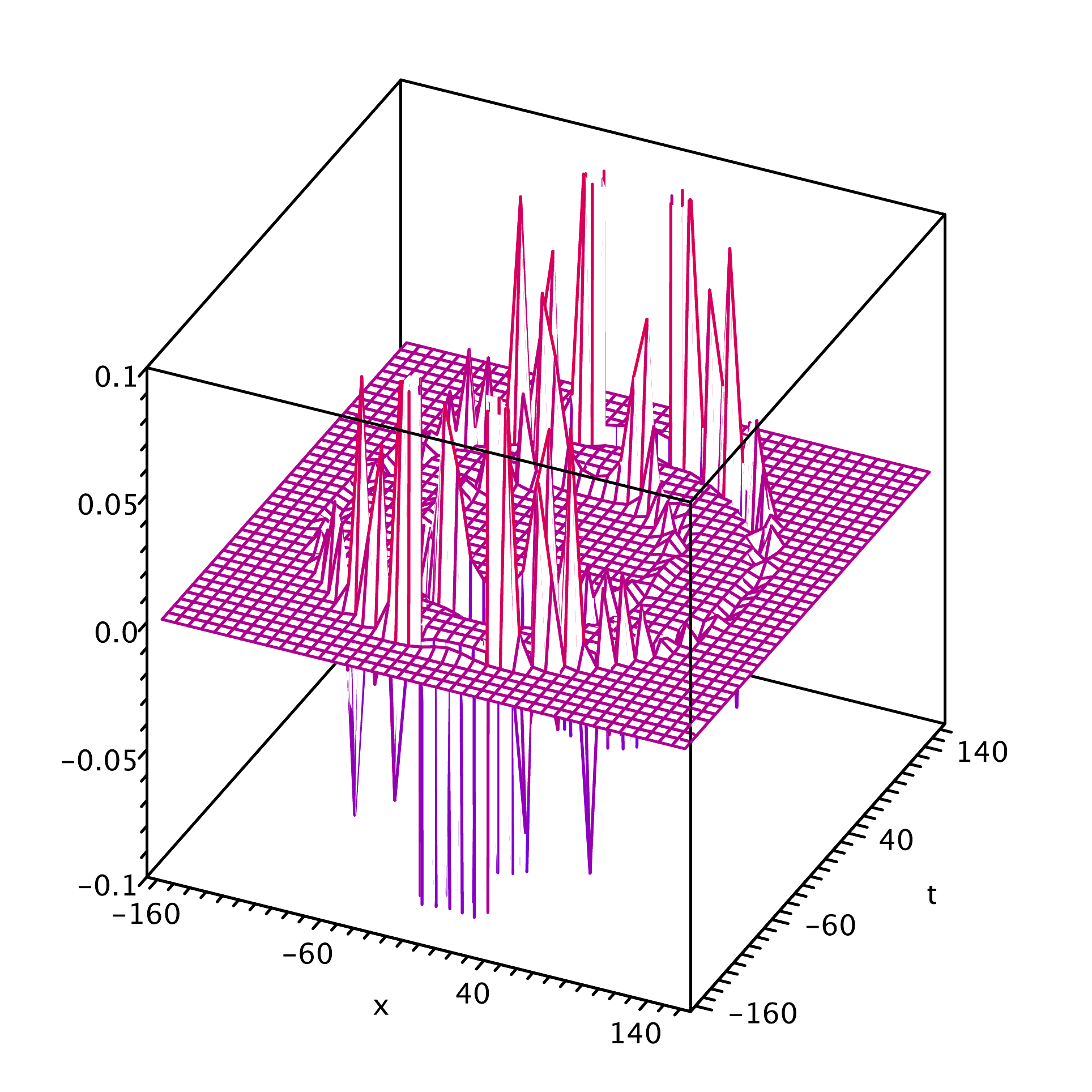}
                             \caption{$G_{zz}$ along the slice $z=30, \; y=0$}
                   \end{subfigure}
                             \quad
 \end{figure*}

The energy density is not  zero everywhere. If we look at cross-sections which do not lay on the $x-t$ axis (across the slice $z=30,\, y=0$ in Fig.  \ref{fig:p7x}), we see a more complicated stress-energy tensor.
  
 \section{Discussion}
 \subsection*{Issues With The TARDIS Metric \label{probs}}
 
Our metric (Eq.\ref{eq:metric}) is symmetric to flipping the sign of the time coordinate or the x,y,z coordinates. As a consequence, observers can travel around CTCs in Fig. \ref{hello} in either the clockwise or counter-clockwise direction. 
 
The light cones which lie on the boundary of the bubble, and transition smoothly between the orientation inside the bubble and the orientation outside the bubble, cannot abide this symmetry. For instance, if the overall metric is symmetric under switching the sign of the x-coordinate, there is no unique and consistent way for the lightcones to tip at the points where the bubble walls intersect the points $x=0$. A broader manifestation of this issue is that there are points where both  $g_{tt}=g_{xx}=0$  and $g_{tx}=g_{xt}=0$.  
 
The issue can be solved by breaking this symmetry, and deciding upon an arrow of time inside and outside the bubble.  This can be done  by adding an extra term to the metric which forces the lightcones at the boundary of the bubble to twist in a deliberate way:

\begin{equation}\begin{split}
 ds^2= & \left[ 1-  h(x,y,z,t)   \left( \frac{2t^2}{x^2 +t^2} \right)  \right](-dt^2 +dx^2) \\
 & +h(x,y,z,t)  \left( \frac{4xt}{x^2 +t^2 } \right)dxdt +dy^2+dz^2 \\
 & +4t^{3} h(x,y,z,t) W(x,y,z,t) dxdt
 \label{eq:metricFIX}
 \end{split} \end{equation}
 where
 \begin{equation} \begin{split}
 W(x,y,z,t)=& \frac{1}{2} \left(  \tanh (\frac{x^{2}-t^{2} (2h(x,y,z,t)-1)+20}{t^2} ) \right.  \\
  & \left.  -\tanh(\frac{x^{2}-t^{2} (2h(x,y,z,t)-1)-20}{t^2})        \right) \;.
 \end{split} \end{equation}
The $W(x,y,z,t)$ function serves to ensure that $g_{tx} \neq 0$ at all points where $g_{tt}=g_{xx}=0$, and also that the additional term vanishes when you are farther from the bubble's boundary.

 \subsection*{Travel Between Arbitrary Points in Spacetime}
While the  geometry as laid out in Eq.\ref{eq:metric} possesses CTCs, we believe that this is only the tip of the iceberg when it comes to using this geometry as a ``time machine." 

Using the Israel-Darmois junction conditions \cite{poisson}, sections of different TARDIS bubble trajectories can be cut out and then stitched together end-to-end. As a consequence, given any two points in Minkowski spacetime, a chain of TARDIS geometries could  be assembled in a way which put the first point in the causal past of the second. 
 
 \subsection*{Questions for Future Consideration}

 \begin{enumerate}
 \item{\emph{Quantum Field Theory}: We can anticipate several interesting questions regarding how quantized fields will behave in the TARDIS geometry. 
 
-Suppose that Israel-Darmois junction conditions are used to smoothly join the  upper half ($t\geq 0$) of the TARDIS metric to the lower half of ($t<0$) a Minkowski metric. Will an electron which enters the bubble from one side emerge from the other side as a positron? 
 
-Since the interior of the bubble looks like Rindler spacetime, will observers travelling within the bubble see Unruh radiation? 

-Since this spacetime necessarily has a Cauchy horizon,  will it possess a mass-inflation type instability \cite{poisson89}? Will some of the arguments of the chronology protection conjecture apply to it?
 
-Should any of the null 3-surfaces associated will this geometry be described as event horizons? If so, will any Hawking radiation be associated with them?
} 
\item{\emph{Dangerous Blueshifting}:  An Alcubierre warp bubble which decelerates from a spacelike trajectory to a timelike trajectory has been shown to blueshift the radiation caught on the bubble surface \cite{McMonigal12}. Since the TARDIS bubbles goes back and forth from being spacelike and timelike, will a similar process take place? }
 
\item{\emph{Singularities}: Our cursory investigation discovered no curvature singularities, and  aside from the issues discussed in Sec.\ref{probs}, the metric is well behaved everywhere. In spite of this, as we discussed in Sec.\ref{causalstru}, it is not unreasonable to expect that this idiosyncratic causal structure should be accompanied by a sort of singularity. Are there any other  singularities in this spacetime? What is their nature?}
\end{enumerate}
\section*{Acknowledgments}
We would like to thank Dr. Chad Galley and  Dr. William Unruh, for their insightful discussions; and Zach Weinersmith for his comments. We would also like to thank Dr. John Smith, for all the Jelly Babies he brought us.
\bibliography{TARDIS6}
\bibliographystyle{plain}

\end{document}